\PassOptionsToPackage{table,xcdraw}{xcolor}
\documentclass[sigconf,nonacm]{acmart}
\setcopyright{none}
\acmISBN{}
\acmDOI{}
\acmPrice{}
\makeatletter
\fancypagestyle{firstpagestyle}{%
  \fancyhf{}
  \fancyhead[L]{2026, Code Watermarking, CLASP}
  \fancyhead[R]{Xu et al.}
  \fancyfoot[C]{\thepage}
}
\fancypagestyle{standardpagestyle}{%
  \fancyhf{}
  \fancyhead[L]{2026, Code Watermarking, CLASP}
  \fancyhead[R]{Xu et al.}
  \fancyfoot[C]{\thepage}
}
\makeatother

\AtBeginDocument{%
  }

\usepackage{graphicx}
\usepackage[utf8]{inputenc} 
\usepackage[T1]{fontenc}    
\usepackage{hyperref}       
\usepackage{url}            
\usepackage{booktabs}       
\usepackage{amsfonts}       
\usepackage{nicefrac}       
\usepackage{microtype}      
\usepackage[skins,listingsutf8]{tcolorbox} 
\usepackage{float}  
\usepackage{amsmath}
\usepackage{array}
\usepackage{multirow}
\usepackage{cellspace}
\usepackage{graphicx}
\usepackage[utf8]{inputenc}
\usepackage[T1]{fontenc}
\usepackage{hyperref}
\usepackage{nicefrac}
\usepackage{listings}
\usepackage{caption}
\usepackage{float}
\usepackage{tcolorbox}
\tcbuselibrary{listingsutf8}
\usepackage{graphicx}
\usepackage{tabularx}
\usepackage{booktabs}
\usepackage{multirow}
\usepackage{wrapfig} 
\usepackage{pifont}
\usepackage{bm}

\usepackage{amssymb}

\definecolor{highlight1}{RGB}{255, 230, 230} 
\definecolor{highlight2}{RGB}{230, 255, 230} 
\definecolor{highlight3}{RGB}{230, 230, 255} 
\definecolor{highlight4}{RGB}{255, 255, 204} 

\newtcbox{\hlcodeA}{on line, boxsep=0pt, left=0pt, right=0pt, top=0pt, bottom=0pt,
  colback=highlight1, colframe=highlight1, boxrule=0pt, sharp corners, before upper=\ttfamily\scriptsize}
\newtcbox{\hlcodeB}{on line, boxsep=0pt, left=0pt, right=0pt, top=0pt, bottom=0pt,
  colback=highlight2, colframe=highlight2, boxrule=0pt, sharp corners, before upper=\ttfamily\scriptsize}
\newtcbox{\hlcodeC}{on line, boxsep=0pt, left=0pt, right=0pt, top=0pt, bottom=0pt,
  colback=highlight3, colframe=highlight3, boxrule=0pt, sharp corners, before upper=\ttfamily\scriptsize}
\newtcbox{\hlcodeD}{on line, boxsep=0pt, left=0pt, right=0pt, top=0pt, bottom=0pt,
  colback=highlight4, colframe=highlight4, boxrule=0pt, sharp corners, before upper=\ttfamily\scriptsize}

\begin{document}

\title{CLASP: Training-Free LLM-Assisted Source Code Watermarking via Semantic-Preserving Transformations}

\author{Rui Xu}
\authornote{Both authors contributed equally to this research.}
\affiliation{%
  \institution{East China Normal University}
  \city{Shanghai}
  \country{China}
}

\author{Jiawei Chen}
\authornotemark[1]
\affiliation{%
  \institution{East China Normal University}
  \city{Shanghai}
  \country{China}
}

\author{Weizhi Liu}
\affiliation{%
  \institution{East China Normal University}
  \city{Shanghai}
  \country{China}
}

\author{Zhaoxia Yin}
\authornote{Corresponding author.}
\affiliation{%
  \institution{East China Normal University}
  \city{Shanghai}
  \country{China}
}

\author{Cong Kong}
\affiliation{%
  \institution{East China Normal University}
  \city{Shanghai}
  \country{China}
}

\author{Xinpeng Zhang}
\affiliation{%
  \institution{Fudan University}
  \city{Shanghai}
  \country{China}
}

\renewcommand{\shortauthors}{Xu et al.}


\begin{abstract}
The proliferation of open-source code and large language models (LLMs) for code generation has amplified the risks of unauthorized reuse and intellectual property infringement.
Source code watermarking offers a potential solution, yet existing methods typically encode watermarks through identifiers, local code patterns, or limited handcrafted edits, leaving them vulnerable to renaming, refactoring, and adaptive watermark removal. 
These limitations hinder the joint achievement of robustness, capacity, generalization, and deployment efficiency.
We propose \textbf{CLASP}, a \textbf{C}ode \textbf{L}LM-\textbf{A}ssisted \textbf{S}emantic-\textbf{P}reserving watermarking framework that enables training-free, plug-and-play watermarking for source code. 
CLASP embeds watermark bits within a fixed space of semantics-preserving transformations, enabling automated watermark insertion with higher capacity while remaining reusable across programming languages and less dependent on brittle lexical features. 
To recover the watermark, CLASP uses reference-code retrieval and differential comparison to identify transformation traces, avoiding task-specific model training while improving robustness to structural edits and adaptive attacks.
Experiments across multiple programming languages show that CLASP consistently outperforms existing baselines in watermark extraction accuracy and robustness, while maintaining code quality under both random removal and adaptive de-watermarking attacks.

\end{abstract}



\maketitle

\section{Introduction}

\begin{table*}[t]
\centering
\caption{
Qualitative comparison of code watermarking methods.
\textbf{Auto} indicates whether the method supports automatic watermark embedding without handcrafted rules.
\textbf{Parser} indicates whether the method avoids parser/AST-dependent processing.
\textbf{Language} indicates whether the method can generalize across programming languages.
\textbf{Plug-and-Play} indicates whether the method can be directly deployed without task-specific retraining or additional adaptation.
\textbf{Capacity} is measured as the maximum number of explicitly decodable bits per function (BPF).
\ding{51}: Supported; \ding{55}: Not supported.
}
\renewcommand{\arraystretch}{0.9}
\small
\setlength{\tabcolsep}{4pt}
\begin{tabular}{lcccccccc}
\toprule
\multirow{2}{*}{\textbf{Method}} 
& \multirow{2}{*}{\textbf{Robustness}} 
& \multicolumn{2}{c}{\textbf{Effectiveness}} 
& \multicolumn{1}{c}{\textbf{Generalization}} 
& \multirow{2}{*}{\textbf{Training-Free}} 
& \multirow{2}{*}{\textbf{Plug-and-Play}} 
& \multirow{2}{*}{\textbf{Trade-off}}
& \multirow{2}{*}{\textbf{Capacity}} \\
\cmidrule(lr){3-4} \cmidrule(lr){5-5}
& & \textbf{Auto} & \textbf{Parser} & \textbf{Language} & & & & \\
\midrule
SrcMarker~\cite{yang2024srcmarker}     & \ding{51} & \ding{55} & \ding{55} & \ding{55} & \ding{55} & \ding{55} & \ding{55} & 8 \\
CodeIP~\cite{guan-etal-2024-codeip}    & \ding{55} & \ding{55} & \ding{51} & \ding{51} & \ding{55} & \ding{55} & \ding{55} & 0 \\
RoSeMary~\cite{zhang2025robust}        & \ding{51} & \ding{55} & \ding{51} & \ding{51} & \ding{55} & \ding{55} & \ding{55} & 0 \\
CodeMark~\cite{li2023towards}          & \ding{51} & \ding{51} & \ding{55} & \ding{55} & \ding{55} & \ding{55} & \ding{55} & 6 \\
\textbf{CLASP (Ours)}                  & \ding{51} & \ding{51} & \ding{51} & \ding{51} & \ding{51} & \ding{51} & \ding{51} & \textbf{32} \\
\bottomrule
\end{tabular}
\label{tab:method_comparison}
\end{table*}

With the rapid growth of the open-source ecosystem, coupled with the strong code generation capabilities of LLMs~\citep{Shrivastava:23,Pei:23}, unauthorized use of source code has become a serious risk~\cite{Khoury:23}. For instance, plagiarists may reuse open-source code with slight modifications and redistribute it under a different license, while LLMs may inadvertently reproduce copyright-protected code from their training data~\cite{sun2022coprotector}. These risks threaten the integrity of the software ecosystem and highlight the need for reliable source code provenance. To address this problem, source code watermarking has emerged as a promising approach~\citep{yang2024srcmarker}. Its goal is to embed recoverable ownership information into code while preserving syntactic correctness, semantic equivalence, and executability~\citep{li2023towards}. The key challenge is to support reliable provenance identification without sacrificing code fidelity or practical usability. Existing methods range from handcrafted semantics-preserving code transformations supported by parser- or abstract syntax tree (AST)-based analysis and rewriting to more recent training-based frameworks~\citep{yang2024srcmarker} that improve watermark naturalness, controllability, or recovery performance.

\begin{figure}[!t] 
  \centering
  \resizebox{0.47\textwidth}{!}{\includegraphics[]{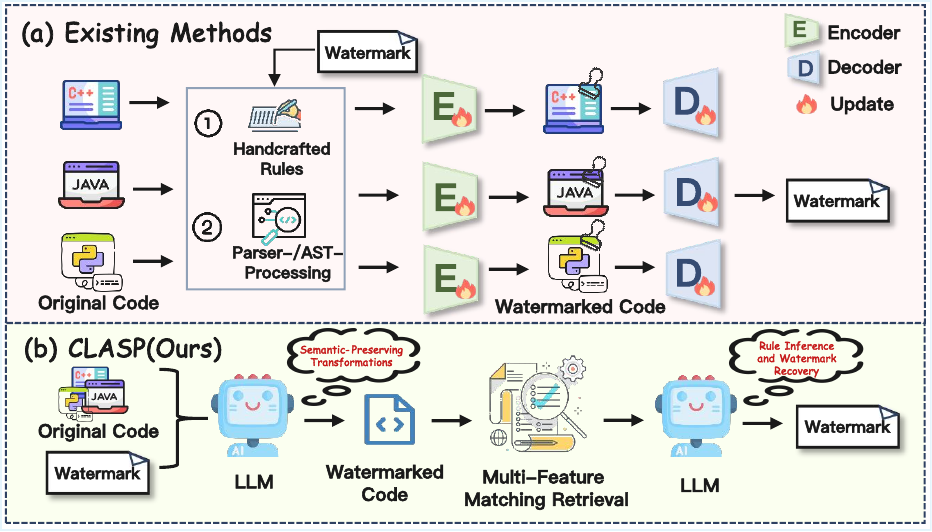}}
  \caption{Comparison between existing source code watermarking methods and our \textbf{CLASP}, which is a training-free and plug-and-play watermarking framework for traceability.}
  \label{fig:intro}
  \vspace{-3ex}
\end{figure}

However, existing methods still face substantial barriers to practical deployment.
As summarized in Table~\ref{tab:method_comparison}, we compare them along seven dimensions following recent discussions on watermarking properties and evaluation criteria~\cite{Zhao:2025:SOK:WatermarkingAIGeneratedContent}. 
As illustrated in Figure~\ref{fig:intro}, in terms of \textit{robustness}, many methods remain vulnerable to variable renaming, code transformation, and adaptive de-watermarking attacks~\citep{li2023towards}. 
Regarding \textit{efficiency}, many still rely on handcrafted rules, parser- or AST-dependent processing, additional training, or task-specific engineering, thereby limiting automation and practical deployment~\citep{yang2024srcmarker,guan-etal-2024-codeip,zhang2025robust,li2023towards}. 
With respect to \textit{generalization}, dependence on language-specific rules or tooling hinders cross-language transfer~\citep{yang2024srcmarker,li2023towards}. 
As for \textit{capacity}, many methods support only limited watermark capacity and cannot scale to higher-capacity designs~\citep{yang2024srcmarker,li2023towards}. 
Beyond these individual drawbacks, many methods also produce watermarked code with syntax errors, compilation failures, or degraded naturalness, resulting in an unsatisfactory \textit{overall trade-off} for practical deployment.

To tackle these challenges, we propose \textbf{CLASP}, a training-free and plug-and-play framework for source code watermarking. As illustrated in Figure~\ref{fig:intro}, unlike prior methods that rely on parser-/AST-dependent rewriting or trained encoder-decoder architectures, CLASP reformulates source code watermarking into two stages: \textit{Semantically Consistent Embedding}, which uses LLMs to perform semantics-aware watermark insertion from a fixed transformation space, and \textit{Differential Comparison Extraction}, which recovers watermark bits through retrieval-grounded comparison against the most likely original code. This design enables automatic watermark embedding, supports cross-language deployment, improves robustness to code modifications during recovery, and avoids the need to train a dedicated decoder. Extensive experiments across multiple programming languages and attack settings show that CLASP consistently outperforms existing baselines while preserving code quality.

The main contributions of this work are as follows:
\begin{itemize}
    \item We propose \textbf{CLASP}, a training-free and plug-and-play source code watermarking framework that combines LLM-assisted semantics-preserving embedding with retrieval-grounded extraction for code traceability and ownership verification across programming languages.
    
    \item We design \textit{Semantically Consistent Embedding}, where prompt-driven LLMs select and apply semantics-preserving transformations from a fixed rule set, enabling automatic and generalizable watermark insertion while preserving functionality, syntax validity, and readability.
    
    \item We design \textit{Differential Comparison Extraction}, which recovers watermark bits by retrieving the most likely original code and inferring transformation traces through code comparison, without training a dedicated decoder.
    
    \item Experiments across multiple programming languages and attack settings show that CLASP outperforms strong baselines in watermark recovery and robustness while preserving code quality and functionality.
\end{itemize}

\begin{figure*}[!t] 
  \centering
  \resizebox{0.8\textwidth}{!}{\includegraphics[]{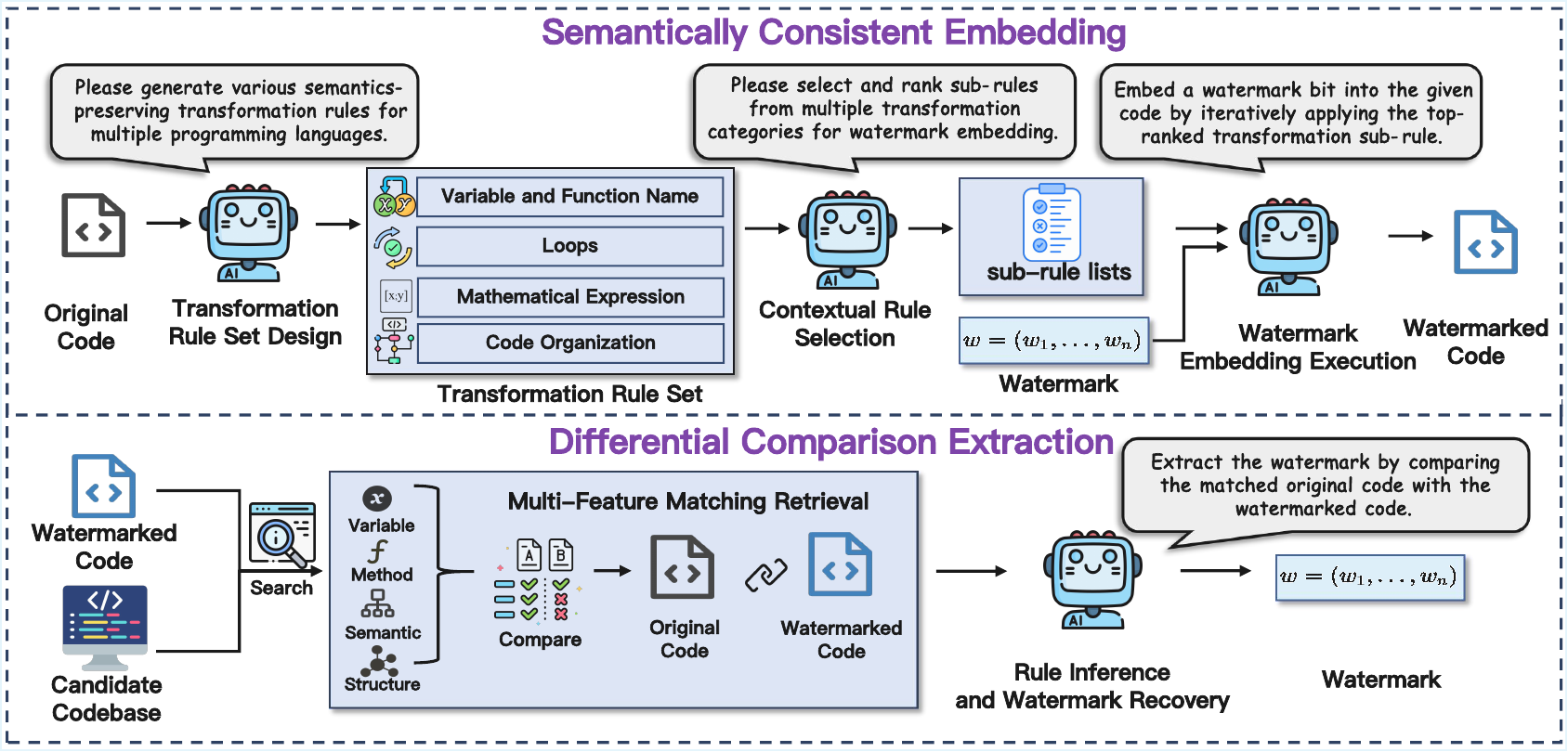}}
  \caption{Overall framework of CLASP. CLASP consists of two stages: \textit{Semantically Consistent Embedding}, which uses an LLM to select context-appropriate transformations from a fixed rule set for watermark embedding, and \textit{Differential Comparison Extraction}, which retrieves the most likely original code candidate and recovers watermark bits through differential comparison.}
  \label{fig:framework}
\end{figure*}

\section{Related Work}
\label{gen_inst}

\textbf{Software watermarking} embeds ownership information into software artifacts for authorship or provenance verification~\cite{Dey:19}. Existing methods are typically divided into static and dynamic watermarking. Static watermarking modifies program structures, binaries, or intermediate representations~\cite{Balachandran:14,Chen:18,Collberg:05,Kang:21,Monden:00}, while dynamic watermarking encodes watermark signals into runtime behaviors or execution states~\cite{Chen:17,Ma:19,Tian:15,Wang:18}. Although effective for compiled programs, these methods are less suitable for raw source code, where syntax, semantics, and usability must be preserved.

\noindent\textbf{Source code watermarking.}
To address this gap, recent work has shifted toward source-level watermarking. Early methods such as RopGen~\cite{Li:22} and NatGen~\cite{Chakraborty:22} rely on semantics-preserving transformations, while later methods introduce learned or hybrid designs. SrcMarker~\cite{yang2024srcmarker} combines rule-based transformations with neural models, ACW~\cite{li2024acw} uses manual transformations in a training-free manner, and CodeMark~\cite{li2023towards} adopts learned variable renaming with AST- and GNN-based modeling. However, existing methods still struggle to balance robustness, generalization, training-free deployment, plug-and-play usability, and high capacity.

\noindent\textbf{LLM-based code transformation.}
Meanwhile, LLMs have shown strong ability in code generation~\cite{OpenAI:22,liu2024deepseek}, transformation, and understanding, and are increasingly used for code completion, translation, and refactoring~\cite{jiang2024survey,zhang2024llm,jana2024cotran,ding2024data}. These advances suggest that LLMs can provide a flexible interface for semantics-preserving code editing without language-specific rewriting pipelines.

Taken together, these lines of research suggest the need for a source code watermarking framework that is more flexible than traditional software watermarking and better balanced than existing source-level methods. Motivated by this, we propose CLASP, which combines LLM-assisted, semantics-preserving embedding with retrieval-grounded watermark extraction in a training-free, plug-and-play framework.

\section{Threat Model}
We consider a source code watermarking scenario with three parties: users, defenders, and adversaries. 

\textit{Capabilities and Goals of Users.}
The user is a code owner who distributes function-level source code to potentially untrusted parties. The user's primary goal is twofold: first, to ensure that the released code remains functionally correct and reusable by recipients; second, to enable post-hoc source traceability and ownership verification in the event of unauthorized reuse or redistribution.

\textit{Capabilities and Goals of Defenders.}
The defender is the watermarking system or its deployer, acting on behalf of the code owner to protect released source code. The defender aims to embed robust, recoverable watermark information into source code before distribution, so that any suspicious reuse or redistribution can be traced to its origin and ownership.

In our setting, the defender starts from the original code $\bm{c}_0 \in \mathcal{C}_0$ and assigns it a watermark $w=(w_1,\dots,w_n)\in\mathcal{W}$, where $n$ denotes the number of watermark bits and $\mathcal{W}$ denotes the set of binary watermark sequences. Using a fixed transformation rule set $\mathcal{T}$, the defender embeds $w$ into $\bm{c}_0$ to obtain the watermarked code $\bm{c}_1 \in \mathcal{C}_1$ before release, while preserving the functionality, syntax validity, and readability of the original code. To support post-hoc verification, the defender also maintains a candidate original codebase $\mathcal{D}$.
Given a suspicious watermarked code sample after distribution, the defender retrieves the closest original candidate $\hat{\bm{c}} \in \mathcal{D}$ and extracts the watermark $\hat{w}$ for verification.

\textit{Capabilities and Goals of Adversaries.}
The adversary is an external party who obtains a copy of the watermarked code and attempts to undermine source traceability or ownership verification. We consider a black-box adversary who does not know the specific transformation rules, the original codebase, the defender-side extraction procedure, or which code regions carry the watermark.

We consider adversaries at two capability levels. ($\mathrm{i}$) \textit{Black-box removal adversary.} The adversary applies indiscriminate functionality-preserving perturbations, such as random variable renaming or arbitrary code transformations, to disrupt potential watermark traces. ($\mathrm{ii}$) \textit{Adaptive de-watermarking adversary.} The adversary employs a general-purpose LLM to paraphrase the watermarked code, attempting to erase watermark signals through semantic-level rewriting while preserving functional correctness. In both settings, the adversary succeeds if the defender's extraction procedure fails to recover the watermark or if ownership verification fails.



\section{Method}
\label{sec:method}

\subsection{Overview}

Source code watermarking faces two key challenges: watermark embedding should be automatic and semantics-preserving across programming languages, while watermark extraction should remain reliable under functionality-preserving code modifications without requiring a dedicated decoder. To address these challenges, we propose CLASP, which embeds watermark bits within a fixed space of semantics-preserving transformations and recovers them via differential comparison with the most likely original code reference. By constraining embedding to context-appropriate, semantics-preserving transformations rather than to unconstrained rewriting, CLASP ensures that the resulting watermarked code preserves functionality, syntactic validity, and naturalness, thereby improving watermark fidelity.

As shown in Figure~\ref{fig:framework}, CLASP consists of two stages: \textit{Semantically Consistent Embedding} and \textit{Differential Comparison Extraction}. In the first stage, CLASP constructs a fixed transformation rule set spanning multiple categories, then uses an LLM to select and iteratively apply context-appropriate sub-rules for each watermark bit to generate watermarked code. In the second stage, given a suspicious watermarked code sample, CLASP retrieves the most likely original candidate from a defender-maintained codebase via multi-feature matching, then compares the two code samples to infer transformation traces and recover watermark bits. By coupling controlled transformation-based embedding with retrieval-grounded extraction, CLASP avoids task-specific decoder training while improving robustness under code modifications and adaptive attacks.

\begin{table*}[t]
\centering
\caption{CLASP's set of transformation rules.}
\renewcommand{\arraystretch}{0.8}
\small
\setlength{\tabcolsep}{4pt}
\resizebox{0.85\textwidth}{!}{
\begin{tabularx}{\textwidth}{
>{\raggedright\arraybackslash}p{3.0cm}
>{\raggedright\arraybackslash}X
!{\vrule width 0.6pt}
>{\raggedright\arraybackslash}p{2.7cm}
>{\raggedright\arraybackslash}X}
\toprule
\multicolumn{2}{c!{\vrule width 0.6pt}}{\textbf{Variable / Function Name}} & \multicolumn{2}{c}{\textbf{Loops}} \\
\midrule
\textbf{Sub-Rule Type} & \textbf{Example (Original $\rightarrow$ Transformed)} 
& \textbf{Sub-Rule Type} & \textbf{Example (Original $\rightarrow$ Transformed)} \\
\midrule
CamelCase to snake\_case & \texttt{testStream()} $\rightarrow$ \texttt{test\_stream()} 
& for to while & \texttt{for(...)} $\rightarrow$ \texttt{while(...)\{...\}} \\
snake\_case to CamelCase & \texttt{my\_var} $\rightarrow$ \texttt{myVar} 
& while to for & \texttt{while(...)} $\rightarrow$ \texttt{for(...)} \\
To PascalCase & \texttt{remove()} $\rightarrow$ \texttt{Remove()} 
& Flatten nested loop & \texttt{for(i)\{...for(j)...\}} $\rightarrow$ \texttt{for(k)...} \\
To UPPERCASE & \texttt{value} $\rightarrow$ \texttt{VALUE} 
& while to do-while & \texttt{while(c)\{...\}} $\rightarrow$ \texttt{do\{...\} while(c);} \\
To lowercase & \texttt{Value} $\rightarrow$ \texttt{value} 
& Step increment & \texttt{i++} $\rightarrow$ \texttt{i+=1} \\
Add suffix & \texttt{data} $\rightarrow$ \texttt{dataVal} 
& Reverse loop & \texttt{for(i=0;...)} $\rightarrow$ \texttt{for(i=n-1;...)} \\
\midrule
\multicolumn{2}{c!{\vrule width 0.6pt}}{\textbf{Math Expression}} & \multicolumn{2}{c}{\textbf{Code Organization}} \\
\midrule
\textbf{Sub-Rule Type} & \textbf{Example (Original $\rightarrow$ Transformed)} 
& \textbf{Sub-Rule Type} & \textbf{Example (Original $\rightarrow$ Transformed)} \\
\midrule
Group ops & \texttt{x + y + z} $\rightarrow$ \texttt{x + (y + z)} 
& Optimize cond. & \texttt{if(x>0) return;} $\rightarrow$ \texttt{if(!(x>0)) return;} \\
Mul to add & \texttt{2 * x} $\rightarrow$ \texttt{x + x} 
& Reorder decl. & \texttt{int a; str b;} $\rightarrow$ \texttt{str b; int a;} \\
Factorization & \texttt{a*b + a*c} $\rightarrow$ \texttt{a*(b + c)} 
& Swap params & \texttt{f(a, b)} $\rightarrow$ \texttt{f(b, a)} \\
Identity transform & \texttt{x*x - y*y} $\rightarrow$ \texttt{(x - y)*(x + y)} 
& Format spacing & \texttt{int x=5;} $\rightarrow$ \texttt{int x = 5;} \\
Div to reciprocal & \texttt{a / b} $\rightarrow$ \texttt{a * (1 / b)} 
& Add braces & \texttt{if(a) return;} $\rightarrow$ \texttt{if(a)\{ return; \}} \\
Pow to mul & \texttt{x*x} $\rightarrow$ \texttt{x * x} 
& Reorder cond. & \texttt{if(a \&\& b)} $\rightarrow$ \texttt{if(b \&\& a)} \\
Expand distributive & \texttt{a*(b + c)} $\rightarrow$ \texttt{a*b + a*c} 
& Insert blank line & \texttt{x=1; y=2;} $\rightarrow$ \texttt{x=1; \textbackslash\textbackslash y=2;} \\
Sqrt-square & \texttt{sqrt(x*x)} $\rightarrow$ \texttt{abs(x)} 
& Adjust op space & \texttt{x=y+z;} $\rightarrow$ \texttt{x = y + z;} \\
& 
& Inline temp var & \texttt{int t=x+y; return t;} $\rightarrow$ \texttt{return x+y;} \\
& 
& Split decl. & \texttt{int x=0,y=1;} $\rightarrow$ \texttt{int x=0; int y=1;} \\
\bottomrule
\end{tabularx}
}
\label{tab:transformation_rules}
\end{table*}

\subsection{Semantically Consistent Embedding}

\textbf{Transformation Rule Set Design.}
CLASP first constructs a fixed set of transformation rules for watermark embedding, providing a reusable space of semantics-preserving transformations for controlled embedding and subsequent recovery. By constraining watermark insertion to this fixed transformation space rather than unconstrained rewriting, CLASP ensures that the resulting watermarked code preserves functionality, syntax validity, and naturalness, thereby improving watermark fidelity. We use LLMs to construct this rule set because they support cross-language code understanding and can identify diverse semantics-preserving transformations under explicit semantic constraints, reducing language-specific bias and improving rule coverage.

Formally, we organize the transformation rule set as
{\setlength{\abovedisplayskip}{2.5pt}
\setlength{\belowdisplayskip}{2.5pt}
\begin{equation}
\mathcal{T} = \bigcup_{i=1}^{m} \mathcal{T}_i, \quad \mathcal{T}_i = \{ T_{i,1},\ T_{i,2},\ \dots,\ T_{i,k_i} \},
\label{eq:rule_set}
\end{equation}
}where \(m=4\) denotes the number of transformation classes, \(\mathcal{T}_i\) the \(i\)-th class, and \(k_i\) the number of sub-rules in that class. Each \(T_{i,j}\) is a specific transformation instance for watermark embedding. The number of transformation classes determines the watermark capacity of CLASP, and higher watermark capacity can be achieved by increasing \(m\). Rule sets for other bit configurations are provided in the Appendix.

The rule set satisfies three properties. First, all transformations are semantics-preserving. Second, the rule set is generated once, fixed for both embedding and extraction, and \textit{kept private by the defender}, so that watermark recovery remains aligned with insertion. Third, the priority order of sub-rules within each class is predefined and fixed, providing a stable basis for subsequent rule selection and bit-wise embedding. Table~\ref{tab:transformation_rules} reports the transformation categories and representative examples for reproducibility, while \textit{the complete deployed rule set and defender-side extraction configuration remain private}. This fixed rule set forms the basis for subsequent bit-wise watermark embedding and extraction.

\noindent\textbf{Contextual Rule Selection with LLM Assistance.}
\label{sec:contextual_rule_selection}
After constructing the fixed transformation rule set, CLASP performs rule selection for watermark embedding. Since programs differ in syntactic structure and semantic constraints, valid transformations must be selected in a context-aware manner to preserve syntax validity, functionality, and readability.
Given the original code \(\bm{c}_0\) and watermark \(w\), CLASP first determines which transformation classes are applicable to \(\bm{c}_0\). For each applicable class \(\mathcal{T}_k\), it scans the sub-rules in predefined priority order and retains the feasible ones to form the candidate list \(\mathcal{L}_k(\bm{c}_0) = [T_{k,1}, T_{k,2}, \dots]\). This design combines the stability of a fixed-rule inventory with the flexibility of context-aware rule selection, thereby improving watermark embedding without language-specific engineering or additional training.

CLASP further introduces a category balancing strategy. For each watermark bit, the LLM first searches for an applicable transformation from previously unused classes to maximize category coverage; only when no new class is applicable does it reuse an existing class. This helps distribute watermark signals more evenly across the transformation space and provides a more stable basis for subsequent extraction.

\noindent\textbf{Watermark Embedding Execution.}
After obtaining the candidate sub-rule lists \(\mathcal{L}_k(\bm{c}_0)\), CLASP embeds the watermark in a bit-wise manner. For each bit, the LLM selects and applies an appropriate transformation from the corresponding candidate list, so that watermark bits are encoded through controlled semantics-preserving transformations rather than unrestricted code rewriting, thereby preserving functionality, syntax validity, and readability.

Given the original code \(\bm{c}_0\), the watermark \(w\), and the candidate sub-rule lists \(\mathcal{L}_k(\bm{c}_0)\), CLASP scans the watermark sequence from left to right. For each bit position \(k\), it selects the transformation \(T_k^*\) from \(\mathcal{L}_k(\bm{c}_0)\). The embedding process is defined as
{\setlength{\abovedisplayskip}{2.5pt}
\setlength{\belowdisplayskip}{2.5pt}
\begin{equation}
\begin{aligned}
\bm{c}^{(k)} &=
\begin{cases}
T_k^*(\bm{c}^{(k-1)}), & \text{if } w_k = 1, \\
\bm{c}^{(k-1)}, & \text{if } w_k = 0,
\end{cases} \\
\bm{c}^{(0)} &= \bm{c}_0,\qquad
\bm{c}_1 = \bm{c}^{(n)}.
\end{aligned}
\label{eq:watermark_embedding}
\end{equation}
}That is, CLASP applies \(T_k^*\) only when \(w_k=1\); otherwise the code remains unchanged.

Because embedding is iterative, earlier transformations may affect later candidate rules. To improve reliability, CLASP uses a bounded retry mechanism: if the current rule is no longer applicable, it scans the remaining sub-rules in \(\mathcal{L}_k(\bm{c}_0)\) until a valid transformation is found or all candidates are exhausted. This improves embedding stability and supports subsequent extraction.

\begin{table}[t]
\centering
\caption{Operational semantic results based on performed operations.
For CSN, \textbf{Pass} denotes the syntax-checking rate; for MBXP/MBPP, \textbf{Pass} denotes the execution-passing rate.
``\textbf{-}'' indicates that the method was not evaluated on the dataset due to prohibitive computational cost. ``\textbf{/}'' indicates that the method cannot be applied to the Python language.}
\renewcommand{\arraystretch}{0.85}
\small
\setlength{\tabcolsep}{4pt}
\resizebox{0.95\linewidth}{!}{
\begin{tabular}{cccccccc}
\toprule
\textbf{Metric} & \textbf{Method} & \textbf{CSN-Java} & \textbf{CSN-JS} & \textbf{MBCPP} & \textbf{MBJP} & \textbf{MBJSP} & \textbf{MBPP} \\
\midrule
\multirow{5}{*}{\textbf{BitAcc (\%) $\uparrow$}}
 & AWT$_\text{code}$~\cite{Abdelnabi:21}   & 93.91 & 89.33 & 97.12 & 93.88 & 83.97 & / \\
 & CALS$_\text{code}$~\cite{Yang:22}  & - & - & 92.89 & 93.31 & 93.50 & / \\
 & SrcMarker~\cite{yang2024srcmarker}           & 97.26 & 96.34 & 96.04 & 99.44 & 96.94 & / \\
 & CodeMark~\cite{li2023towards}            & 89.14 & 87.07 & 68.75 & 72.54 & 64.30 & 70.30 \\
 & \cellcolor[HTML]{E6E6E6}\textbf{CLASP (Ours)}      & \cellcolor[HTML]{E6E6E6}\textbf{98.05} & \cellcolor[HTML]{E6E6E6}\textbf{98.08} & \cellcolor[HTML]{E6E6E6}\textbf{99.64} & \cellcolor[HTML]{E6E6E6}\textbf{99.72} & \cellcolor[HTML]{E6E6E6}\textbf{99.47} & \cellcolor[HTML]{E6E6E6}\textbf{97.85} \\
\midrule
\multirow{5}{*}{\textbf{Pass (\%) $\uparrow$}}
 & AWT$_\text{code}$~\cite{Abdelnabi:21}   & 0.18 & 0.51 & 0.00 & 0.00 & 0.00 & / \\
 & CALS$_\text{code}$~\cite{Yang:22}  & - & - & 68.19 & 68.65 & 76.77 & / \\
 & SrcMarker~\cite{yang2024srcmarker}           & 93.09 & \textbf{100.00} & 97.64 & 97.86 & 97.99 & / \\
 & CodeMark~\cite{li2023towards}            & \textbf{99.87} & 99.97 & 60.47 & 36.22 & 47.80 & 25.77 \\
 & \cellcolor[HTML]{E6E6E6}\textbf{CLASP (Ours)}      & \cellcolor[HTML]{E6E6E6}99.85 & \cellcolor[HTML]{E6E6E6}99.11 & \cellcolor[HTML]{E6E6E6}\textbf{99.35} & \cellcolor[HTML]{E6E6E6}\textbf{99.31} & \cellcolor[HTML]{E6E6E6}\textbf{99.87} & \cellcolor[HTML]{E6E6E6}\textbf{99.69} \\
\bottomrule
\end{tabular}
}
\label{tab:performance}
\vspace{-4ex}
\end{table}

\subsection{Differential Comparison Extraction}
\label{sec:dce}

\textbf{Multi-Feature Matching Retrieval.}
For watermark extraction, CLASP maintains a defender-side candidate codebase \(\mathcal{D}\) and retrieves the most likely original code from it before recovering watermark bits. This design differs from watermark extraction methods that rely on a trained decoder for direct recovery: by grounding extraction in retrieval and code comparison, CLASP remains training-free, is more broadly applicable across programming languages, and is more robust to code modifications and adaptive attacks, while also supporting plug-and-play deployment.

To ensure reliable retrieval, CLASP does not rely on a single matching signal. Since functionality-preserving transformations may alter lexical or structural patterns, retrieval based on only one feature can be unstable. Therefore, CLASP combines multiple complementary similarity signals, including method signature, variable usage, structural features, and normalized semantic form, to identify the most probable original code \(\hat{\bm{c}}\):
{\setlength{\abovedisplayskip}{2.5pt}
\setlength{\belowdisplayskip}{2.5pt}
\begin{equation}
\begin{split}
\hat{\bm{c}} = \arg\max_{\bm{c}_i \in \mathcal{D}} \Big[
& \alpha\, \text{Sim}_m(\bm{c}_i, \bm{c}_1) + \beta\, \text{Sim}_v(\bm{c}_i, \bm{c}_1) \\
& + \gamma\, \text{Sim}_s(\bm{c}_i, \bm{c}_1) + \delta\, \text{Sim}_{\mathrm{sem}}(\bm{c}_i, \bm{c}_1)
\Big],
\end{split}
\label{eq:similarity}
\end{equation}

\begin{equation}
\begin{split}
\text{Sim}_m(\bm{c}_i, \bm{c}_1)
&= 1 -
\frac{
    \text{LevDist}(\mathrm{name}(\bm{c}_i), \mathrm{name}(\bm{c}_1))
}{
    \max\big(|\mathrm{name}(\bm{c}_i)|, |\mathrm{name}(\bm{c}_1)|\big)
},
\end{split}
\end{equation}

\begin{equation}
\text{Sim}_v(\bm{c}_i, \bm{c}_1) =
\frac{|V(\bm{c}_i) \cap V(\bm{c}_1)|}{|V(\bm{c}_i) \cup V(\bm{c}_1)|},
\end{equation}

\begin{equation}
\text{Sim}_s(\bm{c}_i, \bm{c}_1) =
\cos\Big(\vec{f}_s(\bm{c}_i), \vec{f}_s(\bm{c}_1)\Big),
\end{equation}

\begin{equation}
\begin{split}
\text{Sim}_{\mathrm{sem}}(\bm{c}_i, \bm{c}_1)
&= 1 -
\frac{
    \text{LevDist}(\mathrm{norm}(\bm{c}_i), \mathrm{norm}(\bm{c}_1))
}{
    \max\big(|\mathrm{norm}(\bm{c}_i)|, |\mathrm{norm}(\bm{c}_1)|\big)
},
\end{split}
\end{equation}
}where \(\text{Sim}_m\), \(\text{Sim}_v\), \(\text{Sim}_s\), and \(\text{Sim}_{\mathrm{sem}}\) denote similarity scores on method signatures, variable usage, structural features, and semantic features, respectively. Here, \(\mathrm{name}(\cdot)\) denotes the function name, \(V(\cdot)\) is the set of variable identifiers, \(\vec{f}_s(\cdot)\) is a vector of structural token counts, and \(\mathrm{norm}(\cdot)\) is the normalized function string after whitespace removal. In all experiments, we set \(\alpha=\beta=\gamma=\delta=0.25\), selected by grid search as detailed in the Appendix.

Once the most likely original code \(\hat{\bm{c}}\) is identified, CLASP proceeds to infer the applied transformations through differential comparison and then recover the embedded watermark.

\noindent\textbf{Rule Inference and Watermark Recovery.}
After retrieving the most likely original code \(\hat{\bm{c}}\), CLASP uses the LLM to recover the watermark by comparing \(\hat{\bm{c}}\) with the watermarked code \(\bm{c}_1\) and inferring which semantics-preserving transformations have been applied. By using the LLM as the extractor, CLASP keeps the recovery process training-free, plug-and-play, and generalizable across multiple programming languages.

For each watermark bit \(w_k\), CLASP considers the corresponding transformation class \(\mathcal{T}_k\) and scans its candidate sub-rules according to the predefined priority order. Based on the syntactic and semantic consistency between \(\hat{\bm{c}}\) and \(\bm{c}_1\), the LLM identifies the first sub-rule \(T_k^*\) that matches the observed code difference. In this way, each bit is recovered through contextual reasoning over transformation traces rather than direct pattern matching, thereby providing greater robustness to semantics-preserving code modifications.

\begin{figure}[t]
  \centering
  \includegraphics[width=0.95\linewidth]{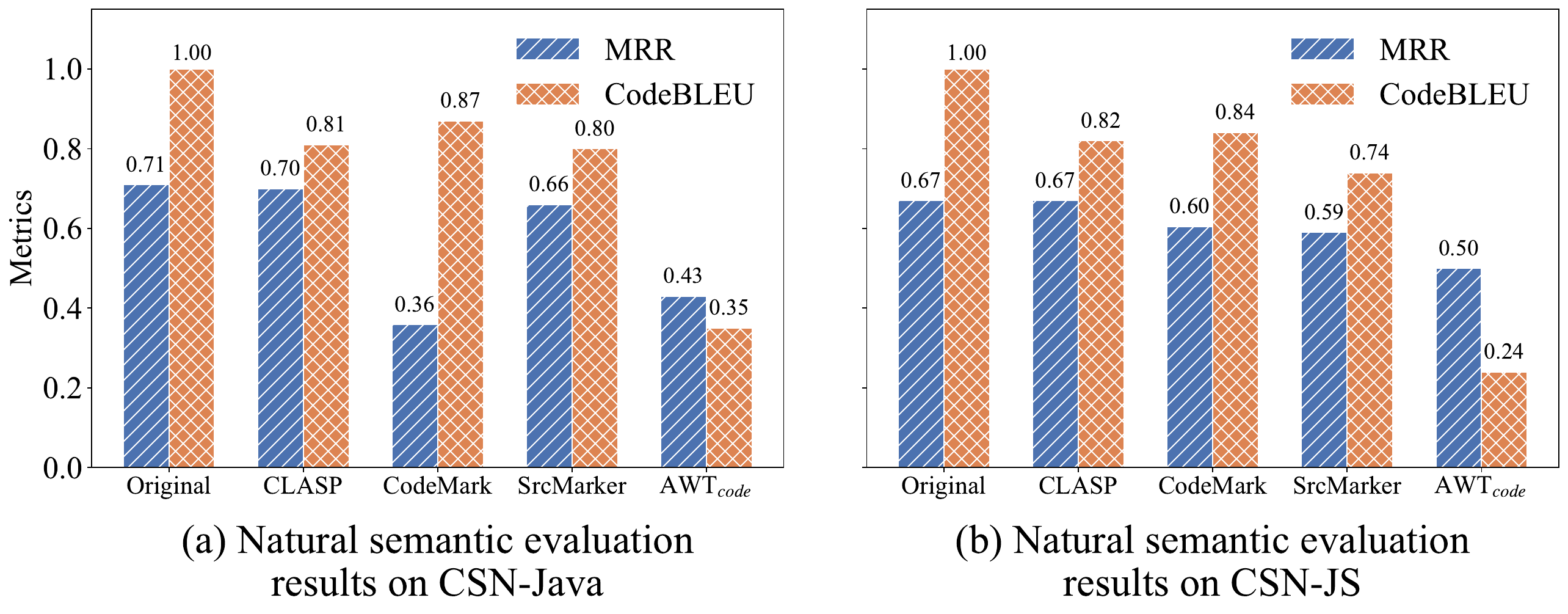} 
  \caption{Natural semantics metrics for CLASP, CodeMark, SrcMarker and AWT$_\text{code}$. "Original" refers to the unwatermarked code.}
  \label{fig:mrr}
\end{figure}

The recovered bit is then defined as
{\setlength{\abovedisplayskip}{3pt}
\setlength{\belowdisplayskip}{3pt}
\begin{equation}
\hat{w}_k =
\begin{cases}
1, & \text{if } T_k^*(\hat{\bm{c}}) \approx \bm{c}_1, \\
0, & \text{otherwise},
\end{cases}
\label{eq:watermark_recovery}
\end{equation}
}where \(\approx\) indicates that applying the candidate sub-rule \(T_k^*\) to the retrieved original code \(\hat{\bm{c}}\) produces code consistent with the watermarked code \(\bm{c}_1\) while preserving syntax validity and functionality.

If the current candidate sub-rule does not yield a valid match, CLASP continues scanning the remaining candidate sub-rules within the same transformation class until a consistent transformation is found or a retry limit is reached. This mechanism further improves the stability of watermark recovery. The final recovered watermark is \(\hat{w} = [\hat{w}_1, \dots, \hat{w}_n]\).

\begin{table*}[t]
\centering
\caption{Performance under black-box removal and adaptive de-watermarking attacks. \textbf{BBR} denotes black-box removal attacks, and \textbf{ADW} denotes adaptive de-watermarking. Here, \textbf{T} denotes random code transformation, \textbf{V} denotes random variable substitution, and \textbf{Pass} is the percentage of attacked code that passes the test cases. ``\textbf{/}'' indicates that the method cannot be applied to the Python language.}
\renewcommand{\arraystretch}{0.8}
\small
\setlength{\tabcolsep}{3.5pt}
\resizebox{0.95\textwidth}{!}{
\begin{tabular}{lllccccccccc}
\toprule
\multirow{2}{*}{\textbf{Dataset}} & \multirow{2}{*}{\textbf{Metric}} & \multirow{2}{*}{\textbf{Method}}
& \multirow{2}{*}{\textbf{No Atk.}}
& \multicolumn{7}{c}{\textbf{BBR}}
& \multicolumn{1}{c}{\textbf{ADW}} \\
\cmidrule(lr){5-11} \cmidrule(lr){12-12}
& & 
& 
& \textbf{T@1} & \textbf{T@2} & \textbf{T@3} 
& \textbf{V@25\%} & \textbf{V@50\%} & \textbf{V@75\%} & \textbf{V@100\%} 
& \textbf{GPT-4o} \\
\midrule

\multirow{9}{*}{\textbf{MBJP}}
& \multirow{3}{*}{\textbf{BitAcc (\%) $\uparrow$}}
& SrcMarker~\cite{yang2024srcmarker} & 99.44 & 86.37 & 82.19 & 78.56 & 81.47 & 77.88 & 67.79 & 56.62 & 50.12 \\
& & CodeMark~\cite{li2023towards} & 72.54 & 68.88 & 67.31 & 64.28 & 67.99 & 62.11 & 57.90 & 49.32 & 48.75 \\
& & \cellcolor[HTML]{E6E6E6}\textbf{CLASP (Ours)} & \cellcolor[HTML]{E6E6E6}\textbf{99.72} & \cellcolor[HTML]{E6E6E6}\textbf{88.84} & \cellcolor[HTML]{E6E6E6}\textbf{86.84} & \cellcolor[HTML]{E6E6E6}\textbf{84.46} & \cellcolor[HTML]{E6E6E6}\textbf{96.03} & \cellcolor[HTML]{E6E6E6}\textbf{95.75} & \cellcolor[HTML]{E6E6E6}\textbf{90.61} & \cellcolor[HTML]{E6E6E6}\textbf{82.29} & \cellcolor[HTML]{E6E6E6}\textbf{76.93} \\
\cmidrule(lr){2-12}
& \multirow{3}{*}{\textbf{Pass (\%) $\uparrow$}}
& SrcMarker~\cite{yang2024srcmarker} & 97.86 & 75.42 & 75.30 & 74.94 & 74.58 & 73.40 & 72.33 & 71.26 & 98.46 \\
& & CodeMark~\cite{li2023towards} & 36.22 & 36.22 & 36.10 & 36.22 & 35.99 & 35.99 & 36.10 & 35.51 & 12.59 \\
& & \cellcolor[HTML]{E6E6E6}\textbf{CLASP (Ours)} & \cellcolor[HTML]{E6E6E6}\textbf{99.31} & \cellcolor[HTML]{E6E6E6}\textbf{93.92} & \cellcolor[HTML]{E6E6E6}\textbf{86.74} & \cellcolor[HTML]{E6E6E6}\textbf{82.32} & \cellcolor[HTML]{E6E6E6}\textbf{91.57} & \cellcolor[HTML]{E6E6E6}\textbf{88.81} & \cellcolor[HTML]{E6E6E6}\textbf{87.97} & \cellcolor[HTML]{E6E6E6}\textbf{87.29} & \cellcolor[HTML]{E6E6E6}\textbf{99.17} \\
\cmidrule(lr){2-12}
& \multirow{3}{*}{\textbf{CodeBLEU (\%) $\uparrow$}}
& SrcMarker~\cite{yang2024srcmarker} & -  & 53.51 & 52.36 & 51.89 & 54.22 & 53.79 & 53.44 & 52.93 & 48.80 \\
& & CodeMark~\cite{li2023towards} & -  & 75.62 & 66.90 & 62.06 & 83.57 & \textbf{78.35} & \textbf{73.12} & 67.14 & \textbf{61.57} \\
& & \cellcolor[HTML]{E6E6E6}\textbf{CLASP (Ours)} & \cellcolor[HTML]{E6E6E6}-  & \cellcolor[HTML]{E6E6E6}\textbf{85.00} & \cellcolor[HTML]{E6E6E6}\textbf{76.74} & \cellcolor[HTML]{E6E6E6}\textbf{70.00} & \cellcolor[HTML]{E6E6E6}\textbf{84.18} & \cellcolor[HTML]{E6E6E6}77.82 & \cellcolor[HTML]{E6E6E6}68.88 & \cellcolor[HTML]{E6E6E6}\textbf{73.45} & \cellcolor[HTML]{E6E6E6}47.85 \\
\midrule

\multirow{9}{*}{\textbf{MBJSP}}
& \multirow{3}{*}{\textbf{BitAcc (\%) $\uparrow$}}
& SrcMarker~\cite{yang2024srcmarker} & 96.94 & 76.47 & 70.64 & 69.70 & 81.93 & 79.33 & 75.63 & 67.06 & 50.72 \\
& & CodeMark~\cite{li2023towards} & 64.30 & 59.47 & 54.27 & 54.77 & 60.79 & 58.34 & 54.74 & 49.28 & 49.94 \\
& & \cellcolor[HTML]{E6E6E6}\textbf{CLASP (Ours)} & \cellcolor[HTML]{E6E6E6}\textbf{99.47} & \cellcolor[HTML]{E6E6E6}\textbf{93.10} & \cellcolor[HTML]{E6E6E6}\textbf{88.08} & \cellcolor[HTML]{E6E6E6}\textbf{88.05} & \cellcolor[HTML]{E6E6E6}\textbf{95.68} & \cellcolor[HTML]{E6E6E6}\textbf{93.62} & \cellcolor[HTML]{E6E6E6}\textbf{90.90} & \cellcolor[HTML]{E6E6E6}\textbf{87.86} & \cellcolor[HTML]{E6E6E6}\textbf{79.02} \\
\cmidrule(lr){2-12}
& \multirow{3}{*}{\textbf{Pass (\%) $\uparrow$}}
& SrcMarker~\cite{yang2024srcmarker} & 97.99 & 96.86 & 97.24 & 96.61 & 96.24 & \textbf{96.36} & \textbf{95.23} & \textbf{94.35} & 99.25 \\
& & CodeMark~\cite{li2023towards} & 47.80 & 46.80 & 46.68 & 46.68 & 46.93 & 46.68 & 46.68 & 46.55 & 0.50 \\
& & \cellcolor[HTML]{E6E6E6}\textbf{CLASP (Ours)} & \cellcolor[HTML]{E6E6E6}\textbf{99.87} & \cellcolor[HTML]{E6E6E6}\textbf{98.37} & \cellcolor[HTML]{E6E6E6}\textbf{98.12} & \cellcolor[HTML]{E6E6E6}\textbf{96.99} & \cellcolor[HTML]{E6E6E6}\textbf{96.49} & \cellcolor[HTML]{E6E6E6}93.60 & \cellcolor[HTML]{E6E6E6}90.97 & \cellcolor[HTML]{E6E6E6}89.34 & \cellcolor[HTML]{E6E6E6}\textbf{99.87} \\
\cmidrule(lr){2-12}
& \multirow{3}{*}{\textbf{CodeBLEU (\%) $\uparrow$}}
& SrcMarker~\cite{yang2024srcmarker} & -  & 49.24 & 48.72 & 48.54 & 49.31 & 49.10 & 48.91 & 48.43 & 49.80 \\
& & CodeMark~\cite{li2023towards} & -  & 66.12 & 58.80 & 55.08 & \textbf{76.09} & \textbf{72.07} & \textbf{68.26} & \textbf{62.80} & \textbf{54.47} \\
& & \cellcolor[HTML]{E6E6E6}\textbf{CLASP (Ours)} & \cellcolor[HTML]{E6E6E6}\cellcolor[HTML]{E6E6E6}- & \cellcolor[HTML]{E6E6E6}\textbf{75.03} & \cellcolor[HTML]{E6E6E6}\textbf{65.90} & \cellcolor[HTML]{E6E6E6}\textbf{60.06} & \cellcolor[HTML]{E6E6E6}72.85 & \cellcolor[HTML]{E6E6E6}64.47 & \cellcolor[HTML]{E6E6E6}60.27 & \cellcolor[HTML]{E6E6E6}57.02 & \cellcolor[HTML]{E6E6E6}48.55 \\
\midrule

\multirow{9}{*}{\textbf{MBCPP}}
& \multirow{3}{*}{\textbf{BitAcc (\%) $\uparrow$}}
& SrcMarker~\cite{yang2024srcmarker} & 96.04 & 85.21 & 77.65 & 73.56 & 83.05 & 79.52 & 72.64 & 60.18 & 49.18 \\
& & CodeMark~\cite{li2023towards} & 68.75 & 64.23 & 61.85 & 60.11 & 63.94 & 58.84 & 57.04 & 49.51 & 49.15 \\
& & \cellcolor[HTML]{E6E6E6}\textbf{CLASP (Ours)} & \cellcolor[HTML]{E6E6E6}\textbf{99.64} & \cellcolor[HTML]{E6E6E6}\textbf{92.64} & \cellcolor[HTML]{E6E6E6}\textbf{91.36} & \cellcolor[HTML]{E6E6E6}\textbf{89.20} & \cellcolor[HTML]{E6E6E6}\textbf{94.93} & \cellcolor[HTML]{E6E6E6}\textbf{92.67} & \cellcolor[HTML]{E6E6E6}\textbf{91.66} & \cellcolor[HTML]{E6E6E6}\textbf{90.35} & \cellcolor[HTML]{E6E6E6}\textbf{83.48} \\
\cmidrule(lr){2-12}
& \multirow{3}{*}{\textbf{Pass (\%) $\uparrow$}}
& SrcMarker~\cite{yang2024srcmarker} & 97.64 & 79.84 & 79.45 & 79.71 & 78.66 & 77.62 & 74.48 & 73.95 & \textbf{99.74} \\
& & CodeMark~\cite{li2023towards} & 60.47 & 47.25 & 46.73 & 46.99 & 46.34 & 46.86 & 46.60 & 45.42 & 2.09 \\
& & \cellcolor[HTML]{E6E6E6}\textbf{CLASP (Ours)} & \cellcolor[HTML]{E6E6E6}\textbf{99.35} & \cellcolor[HTML]{E6E6E6}\textbf{95.16} & \cellcolor[HTML]{E6E6E6}\textbf{92.15} & \cellcolor[HTML]{E6E6E6}\textbf{91.23} & \cellcolor[HTML]{E6E6E6}\textbf{90.18} & \cellcolor[HTML]{E6E6E6}\textbf{88.48} & \cellcolor[HTML]{E6E6E6}\textbf{88.09} & \cellcolor[HTML]{E6E6E6}\textbf{86.91} & \cellcolor[HTML]{E6E6E6}\textbf{99.74} \\
\cmidrule(lr){2-12}
& \multirow{3}{*}{\textbf{CodeBLEU (\%) $\uparrow$}}
& SrcMarker~\cite{yang2024srcmarker} & - & 44.88 & 44.86 & 44.99 & 44.08 & 43.81 & 43.57 & 42.98 & 50.75 \\
& & CodeMark~\cite{li2023towards} & -  & \textbf{69.60} & 59.05 & 52.72 & \textbf{78.55} & \textbf{72.64} & \textbf{66.94} & \textbf{59.67} & \textbf{52.04} \\
& & \cellcolor[HTML]{E6E6E6}\textbf{CLASP (Ours)} & \cellcolor[HTML]{E6E6E6}- & \cellcolor[HTML]{E6E6E6}66.65 & \cellcolor[HTML]{E6E6E6}\textbf{59.05} & \cellcolor[HTML]{E6E6E6}\textbf{54.49} & \cellcolor[HTML]{E6E6E6}71.21 & \cellcolor[HTML]{E6E6E6}64.64 & \cellcolor[HTML]{E6E6E6}61.48 & \cellcolor[HTML]{E6E6E6}58.25 & \cellcolor[HTML]{E6E6E6}4.88 \\
\midrule

\multirow{6}{*}{\textbf{MBPP}}
& \multirow{2}{*}{\textbf{BitAcc (\%) $\uparrow$}}
& 
SrcMarker~\cite{yang2024srcmarker} & / & / & / & / & / & / & / & / & / \\
& & 
CodeMark~\cite{li2023towards} & 70.30 & 50.62 & 50.46 & 50.44 & 49.46 & 49.02 & 49.72 & 49.92 & 50.18 \\
& & \cellcolor[HTML]{E6E6E6}\textbf{CLASP (Ours)} & \cellcolor[HTML]{E6E6E6}\textbf{97.85} & \cellcolor[HTML]{E6E6E6}\textbf{93.85} & \cellcolor[HTML]{E6E6E6}\textbf{92.96} & \cellcolor[HTML]{E6E6E6}\textbf{90.80} & \cellcolor[HTML]{E6E6E6}\textbf{89.13} & \cellcolor[HTML]{E6E6E6}\textbf{95.82} & \cellcolor[HTML]{E6E6E6}\textbf{94.16} & \cellcolor[HTML]{E6E6E6}\textbf{92.82} & \cellcolor[HTML]{E6E6E6}\textbf{79.72} \\
\cmidrule(lr){2-12}
& \multirow{2}{*}{\textbf{Pass (\%) $\uparrow$}}
& 
SrcMarker~\cite{yang2024srcmarker} & / & / & / & / & / & / & / & / & / \\
& & 
CodeMark~\cite{li2023towards} & 25.77 & 25.70 & 25.66 & 25.54 & 25.36 & 25.67 & 25.46 & 25.26 & 7.80 \\
& & \cellcolor[HTML]{E6E6E6}\textbf{CLASP (Ours)} & \cellcolor[HTML]{E6E6E6}\textbf{99.69} & \cellcolor[HTML]{E6E6E6}\textbf{95.64} & \cellcolor[HTML]{E6E6E6}\textbf{95.43} & \cellcolor[HTML]{E6E6E6}\textbf{95.22} & \cellcolor[HTML]{E6E6E6}\textbf{95.11} & \cellcolor[HTML]{E6E6E6}\textbf{95.85} & \cellcolor[HTML]{E6E6E6}\textbf{91.70} & \cellcolor[HTML]{E6E6E6}\textbf{88.28} & \cellcolor[HTML]{E6E6E6}\textbf{99.38} \\
\cmidrule(lr){2-12}
& \multirow{2}{*}{\textbf{CodeBLEU (\%) $\uparrow$}}
& 
SrcMarker~\cite{yang2024srcmarker} & / & / & / & / & / & / & / & / & / \\
& & 
CodeMark~\cite{li2023towards} & -  & 60.45 & 57.08 & 52.30 & 60.45 & 57.08 & 52.30 & 47.04 & 0.45 \\
& & \cellcolor[HTML]{E6E6E6}\textbf{CLASP (Ours)} & \cellcolor[HTML]{E6E6E6}-  & \cellcolor[HTML]{E6E6E6}\textbf{64.01} & \cellcolor[HTML]{E6E6E6}\textbf{61.10} & \cellcolor[HTML]{E6E6E6}\textbf{57.14} & \cellcolor[HTML]{E6E6E6}\textbf{52.10} & \cellcolor[HTML]{E6E6E6}\textbf{65.22} & \cellcolor[HTML]{E6E6E6}\textbf{68.28} & \cellcolor[HTML]{E6E6E6}\textbf{67.44} & \cellcolor[HTML]{E6E6E6}\textbf{41.19} \\
\bottomrule
\end{tabular}
}
\label{tab:robustness_attack_results}
\end{table*}

\section{Evaluation}
\label{others}
In this section, we evaluate CLASP. We first describe the experimental setup in Section~\ref{sec:setup}. For CLASP,  we evaluate its watermark accuracy and fidelity (Section~\ref{sec:accuracy}), robustness (Section~\ref{sec:robustness}), effectiveness (Section~\ref{sec:Effectiveness}), capacity (Section~\ref{sec:Cpacity}), and ablation study (Section~\ref{sec:Ablation Study}).

\subsection{Experiment Setup}
\label{sec:setup}
\textbf{Datasets and Preprocessing.} To evaluate CLASP across languages, we use CodeSearchNet and MBXP benchmarks covering C, C++, Java, JavaScript, and Python. For semantic evaluation, we use CSN-Java and CSN-JS~\cite{Husain:19}, in which functions are paired with natural-language descriptions from open-source projects. For execution-based validation, we use MBXP~\cite{Athiwaratkun:22}, including MBCPP, MBJP, MBJSP, and MBPP. Each function embeds a 4-bit watermark. Detailed settings and results on GitHub subsets are provided in Appendix.

\noindent\textbf{Baselines and LLMs.} We choose CodeMark, SrcMarker, AWT$_\text{code}$, and CALS$_\text{code}$ proposed in SrcMarker as baselines. AWT$_\text{code}$ and CALS$_\text{code}$ are adapted from AWT~\cite{Abdelnabi:21} and CALS~\cite{Yang:22}, both originally developed for natural language watermarking. AWT$_\text{code}$ keeps the AWT architecture but is trained on source code datasets, while CALS$_\text{code}$ replaces BERT~\cite{Devlin:19} with CodeBERT~\cite{Feng:20} to better fit source code data. For CLASP, we use GPT-4o for embedding and extraction; results with DeepSeek-V3 and Gemini 1.5 Pro are reported in the Appendix.

\subsection{Watermark Accuracy and Fidelity}
\label{sec:accuracy}  

\textbf{Metrics.} We evaluate watermarking performance from two perspectives: \textbf{accuracy} and \textbf{fidelity}. 
For watermark accuracy, we use Bit Accuracy (BitAcc), the percentage of watermark bits correctly extracted. 
For watermark fidelity, we evaluate whether the watermarked code preserves its operational correctness after transformation. 
Specifically, for the CSN datasets, we use syntax checking, since the functions are not independently compilable; following prior work, we adopt \texttt{tree-sitter} to detect AST errors. 
For the MBXP datasets, we use the pass rate, defined as the fraction of watermarked programs that pass all unit tests. 
In addition, following prior studies, we assess natural semantic fidelity using CodeBLEU~\cite{ren2020codebleu} and MRR. 
CodeBLEU measures code similarity using syntax, data flow, and n-gram matching, while MRR evaluates whether watermarking affects the rank of code retrieval for natural language queries, with a higher score indicating better semantic retention. 
For MRR computation, we use a fine-tuned CodeBERT~\cite{Feng:20}.

\noindent\textbf{Accuracy results.} Table~\ref{tab:performance} shows that CLASP achieves the best overall watermark extraction accuracy across the evaluated datasets. Compared with CodeMark, SrcMarker, AWT$_\text{code}$, and CALS$_\text{code}$, CLASP consistently obtains the highest BitAcc, demonstrating stronger watermark recovery effectiveness across programming languages. On the two CSN datasets, CLASP outperforms all baselines by clear margins. On the MBXP benchmarks, its advantage is even more pronounced, where it consistently achieves near-perfect BitAcc while several baselines show noticeable degradation.

\begin{table}[t]
\centering
\caption{Comparison of runtime and per-sample economic cost across different watermarking methods.}
\small
\setlength{\tabcolsep}{2pt}
\renewcommand{\arraystretch}{0.95}
\resizebox{0.95\linewidth}{!}{%
\begin{tabular}{lcccc}
\toprule
\textbf{Method} & AWT$_\text{code}$~\cite{Abdelnabi:21} & SrcMarker~\cite{yang2024srcmarker} & CodeMark~\cite{li2023towards} & \cellcolor[HTML]{E6E6E6}\textbf{CLASP (Ours)} \\
\midrule
BitAcc (\%) $\uparrow$           & 93.91   & 97.26           & 89.14           & \cellcolor[HTML]{E6E6E6}\textbf{98.05} \\
Pass (\%) $\uparrow$             & 0.18    & 93.09           & \textbf{99.87}  & \cellcolor[HTML]{E6E6E6}99.85 \\
Training Time (h) $\downarrow$   & 61.50   & 13.32           & 13.87           & \cellcolor[HTML]{E6E6E6}\textbf{0} \\
Embedding Time (s) $\downarrow$  & 0.1055  & 0.0741          & \textbf{0.0387} & \cellcolor[HTML]{E6E6E6}3.3333 \\
Extraction Time (s) $\downarrow$ & \textbf{0.0023} & 0.0034 & 0.0074          & \cellcolor[HTML]{E6E6E6}1.3333 \\
Total Time (h) $\downarrow$      & 61.50   & 13.32           & 13.87           & \cellcolor[HTML]{E6E6E6}\textbf{5.85} \\
Economic Cost (\$) $\downarrow$  & 0.0123  & 0.0027          & 0.0029          & \cellcolor[HTML]{E6E6E6}\textbf{0.0020} \\
\bottomrule
\end{tabular}%
}
\label{tab:economic cost}
\vspace{-3ex}
\end{table}

\noindent\textbf{Operational semantic results.} The operational semantic results are shown in Table~\ref{tab:performance}. AWT$_\text{code}$ performs poorly in execution-based evaluation, with nearly zero pass rates on MBXP datasets, indicating that it often fails to preserve executable functionality after watermark embedding. CALS$_\text{code}$ improves over AWT$_\text{code}$, but still fails a substantial portion of test cases. SrcMarker achieves relatively strong operational fidelity, but still suffers occasional syntax or execution failures. Although CodeMark maintains high pass rates on the CSN datasets, its execution pass rates drop sharply on MBXP benchmarks, suggesting that it struggles to preserve actual runtime behavior. In contrast, CLASP consistently achieves near-perfect pass rates across datasets, demonstrating that its semantics-preserving embedding strategy maintains code validity and functionality while enabling accurate watermark recovery.

\noindent\textbf{Natural semantic results.} Figure~\ref{fig:mrr} shows the results on natural semantic fidelity. Higher MRR and CodeBLEU indicate better preservation of naturalness and semantic consistency after watermark embedding. Overall, CLASP remains closest to the original code and achieves the best balance between MRR and CodeBLEU on both datasets, showing stronger natural semantic fidelity than the baselines. In contrast, SrcMarker and AWT$_\text{code}$ lead to more noticeable degradation in naturalness, while CodeMark presents a different trade-off: although it preserves relatively high CodeBLEU, its MRR drops more substantially, suggesting that it is less effective at maintaining retrieval-based natural semantic consistency.

\subsection{Robustness}
\label{sec:robustness}

\textbf{Random removal attack.}  
To evaluate the \textbf{robustness} of CLASP, we consider a black-box removal adversary that knows code transformations may be used for watermark embedding, but does not know the exact transformation rules or extraction procedure. The attacker performs random functionality-preserving edits by renaming \(25\%\), \(50\%\), \(75\%\), and \(100\%\) of variables, and applying up to 1, 2, or 3 random code transformations to each program. As shown in Table~\ref{tab:robustness_attack_results}, the BitAcc of all methods decreases as the attack intensity increases. Nevertheless, CLASP consistently maintains higher BitAcc than the baselines across all MBXP datasets under both transformation-based and identifier-level attacks, while also preserving high pass rates after attack. This advantage is especially clear under variable-substitution attacks, where CLASP remains much more stable than SrcMarker and CodeMark as the renaming ratio increases. Because CLASP introduces a category balancing strategy during embedding, it does not rely on a single transformation pattern to carry watermark signals. As a result, even if random transformation attacks destroy watermark evidence in some code regions, the remaining transformation categories can still support accurate recovery. This makes CLASP more robust to functionality-preserving random edits while maintaining code executability.

\noindent\textbf{Adaptive de-watermarking.}  
We further evaluate CLASP under a stronger adaptive adversary that is aware of the general LLM-based embedding setting, but does not know the exact watermarking strategy. In this setting, the attacker uses GPT-4o to rewrite the watermarked code while preserving its intended functionality, aiming to remove watermark signals through higher-level paraphrasing and reorganization. As shown in Table~\ref{tab:robustness_attack_results}, CLASP consistently achieves the highest BitAcc across MBXP datasets under this stronger attack setting, while also maintaining high pass rates after rewriting. By contrast, the baselines show much sharper degradation, with several BitAcc results approaching random-guessing level. This is because adaptive rewriting changes not only identifiers and local syntax, but also higher-level expression patterns and code organization, making methods that rely more heavily on surface-form signals easier to weaken. By contrast, CLASP performs extraction through differential comparison with the most likely original code candidate, allowing it to recover watermark bits from persistent transformation-level differences rather than exact lexical matches, thereby maintaining high watermark recoverability and code executability under adaptive de-watermarking attacks.


\subsection{Effectiveness}
\label{sec:Effectiveness}  

\subsubsection{Economic Cost}
\textbf{Metrics.} We compare CLASP with the baselines using per-sample cost, including total time (training, embedding, and extraction time) and economic cost. In addition, to highlight the trade-off between efficiency and watermark quality, we also report Bit Accuracy (BitAcc) and Pass rate, where Pass denotes the syntax-checking rate on the CSN-Java dataset. SrcMarker, AWT\(_\text{code}\), and CodeMark require non-trivial computational resources for model training, so we estimate their training costs using a common market rental price of \$2/hour for GPU usage. Due to the extremely high runtime of CALS\(_\text{code}\) (over 342 hours), we do not include it in the comparison.

\begin{table}[t]
  \centering
  \small
  \setlength{\tabcolsep}{2pt}
  \caption{Plug-and-play evaluation of CLASP under cross-LLM embedding and extraction, measured by Bit Accuracy (BitAcc) across MBXP datasets.}
  \resizebox{0.95\linewidth}{!}{%
  \begin{tabular}{lcccccc}
    \toprule
    \textbf{Watermarker} & \textbf{Extractor} & \textbf{MBJP (\%)} & \textbf{MBJSP (\%)} & \textbf{MBCPP (\%)} & \textbf{MBPP (\%)} \\
    \midrule
    \multirow{3}{*}{GPT-4o} & GPT-4o & \textbf{99.72} & 99.47 & \textbf{99.64} & \textbf{97.85} \\
    & DeepSeek-V3 & 99.30 & \textbf{99.72} & 98.85 & 95.51 \\
    & Gemini-1.5-pro & 99.26 & \textbf{99.72} & 99.48 & 95.38 \\
    \midrule
    \multirow{3}{*}{DeepSeek-V3} & DeepSeek-V3 & 98.69 & \textbf{99.56} & \textbf{98.75} & 98.37 \\
    & GPT-4o & 95.59 & 97.62 & 98.28 & 96.73 \\
    & Gemini-1.5-pro & \textbf{99.62} & 95.67 & 98.28 & \textbf{98.99} \\
    \midrule
    \multirow{3}{*}{Gemini-1.5-pro} & Gemini-1.5-pro & 98.70 & 98.19 & 98.13 & \textbf{97.37} \\
    & GPT-4o & 98.79 & 98.43 & 98.54 & 96.38 \\
    & DeepSeek-V3 & \textbf{99.34} & \textbf{99.75} & \textbf{98.98} & 96.96 \\
    \bottomrule
  \end{tabular}%
  }
  \label{tab:cross-llm-performance}
  \vspace{-2ex}
\end{table}

\noindent\textbf{Results.} Table~\ref{tab:economic cost} shows a clear trade-off between inference efficiency and watermark quality. Although CLASP requires longer embedding and extraction time than the baselines at inference time, it achieves the highest BitAcc and near-perfect syntax pass rate, demonstrating substantially better plug-and-play watermarking quality. In contrast, methods such as AWT\(_\text{code}\), SrcMarker, and CodeMark are faster once trained, but their performance is either less accurate or less stable in terms of syntax correctness. At the same time, because CLASP avoids any upfront training costs, it still achieves the lowest overall cost and the shortest end-to-end runtime among the methods compared. These results show that CLASP offers a favorable practical trade-off: it sacrifices some inference speed to obtain much stronger watermark extraction accuracy and syntax preservation, while remaining more economical overall than training-based baselines.

\begin{figure}[t]
  \centering
  \includegraphics[width=0.95\linewidth]{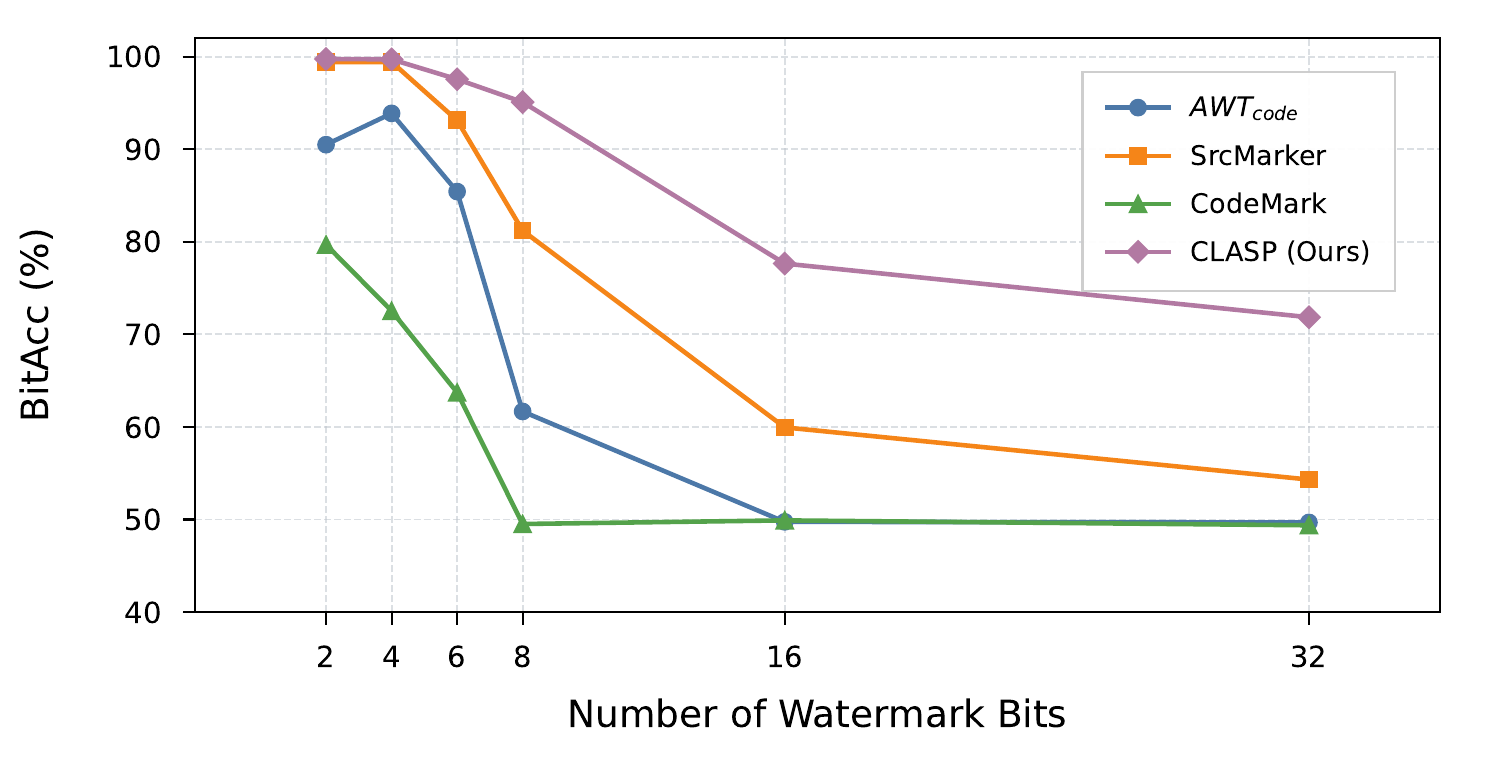} 
  \caption{BitAcc changes with respect to watermark capacity (number of
watermark bits embedded into a single function).}
  \label{fig:capacity}
  \vspace{-3ex}
\end{figure}

\subsubsection{Generalization}
\label{sec:Plug-and-Play}
\textbf{Metrics.} To evaluate the plug-and-play capability of CLASP, we conduct cross-LLM experiments in which watermark embedding and extraction are performed by different LLMs. Specifically, we instantiate the watermarker and extractor with three representative LLMs, namely \textbf{GPT-4o}, \textbf{DeepSeek-V3}, and \textbf{Gemini 1.5 Pro}, and evaluate all cross-model combinations. We use Bit Accuracy (BitAcc) as the metric, which measures the percentage of watermark bits correctly recovered. A method with strong plug-and-play capability should maintain high extraction accuracy even when the embedding and extraction stages are carried out by different LLMs.

\noindent\textbf{Results.} 
Table~\ref{tab:cross-llm-performance} shows that CLASP maintains consistently high BitAcc across different watermarker--extractor combinations, indicating that it does not rely on a fixed model pairing and remains stable across heterogeneous LLM settings. This is because both embedding and extraction are grounded in the same fixed transformation space, rather than tied to a model-specific decoder or parameterization. Moreover, as shown in Table~\ref{tab:performance}, several baseline methods exhibit limited language generalization due to their reliance on language-specific parsing and transformation mechanisms, and thus cannot be applied to Python or C programs, making them non-plug-and-play. In contrast, CLASP supports multiple programming languages while maintaining strong watermark recovery and execution fidelity. These results demonstrate strong cross-language generalization and plug-and-play capability.

\subsection{Capacity}
\label{sec:Cpacity}
We conduct capacity experiments on the MBJP dataset and compare the BitAcc of different methods across watermark capacities of 2, 4, 6, 8, 16, and 32 bits. As shown in Figure~\ref{fig:capacity}, the BitAcc of all methods decreases as the number of embedded bits increases, indicating that reliable watermark embedding and extraction become more difficult at higher capacities. Notably, after 8 bits, the other three methods all drop to near-random performance, suggesting that their watermark embedding and extraction fail under high-capacity settings. In contrast, CLASP maintains high BitAcc not only at low-bit settings but also under 16-bit and 32-bit watermarking. These results show that CLASP supports higher watermark capacity while preserving reliable recovery performance.

\subsection{Ablation Study}
\label{sec:Ablation Study}

\textbf{Metrics.}
We conduct ablation experiments on MBJP to evaluate the necessity of the retrieval, embedding, and extraction components in CLASP. For retrieval, we compare the full multi-feature retrieval of CLASP with several single-feature variants, including name-only, variable-only, structure-only, and semantic-only retrieval, as well as the original-code setting. For embedding, we remove the fixed transformation rule set. For extraction, we replace retrieval-grounded differential comparison with direct bit prediction without using the original code. We report Top-1 and Top-5 for retrieval variants, and BitAcc for final watermark recovery.

\noindent\textbf{Results.}
Table~\ref{tab:ablation} reports the ablation results. For retrieval, the full multi-feature retrieval achieves the same BitAcc as using the original code, whereas all single-feature variants perform worse, indicating that combining complementary features is necessary to reliably identify the original reference and support downstream extraction. For embedding, removing the fixed transformation rule set results in a sharp drop in BitAcc, indicating that the shared transformation space is necessary to align watermark insertion with subsequent recovery. For extraction, replacing retrieval-grounded differential comparison with direct bit prediction without using the original code yields near-random performance, indicating that accurate watermark recovery relies on \textit{Differential Comparison Extraction}. Overall, these results confirm the necessity of retrieval, fixed-rule embedding, and differential comparison in CLASP.

\begin{table}[t]
\centering
\caption{Ablation results on MBJP. ``Multi-feature retrieval'' denotes the full retrieval setting of CLASP, and ``--'' indicates that retrieval accuracy is not applicable.}
\label{tab:ablation}
\resizebox{0.95\linewidth}{!}{
\begin{tabular}{llccc}
\toprule
\textbf{Component} & \textbf{Variant} & \textbf{Top-1 (\%)} $\uparrow$ & \textbf{Top-5 (\%)} $\uparrow$ & \textbf{BitAcc (\%)} $\uparrow$ \\
\midrule
\multirow{6}{*}{Retrieval}
& Original code & 100.00 & 100.00 & 99.72 \\
& Multi-feature retrieval (full CLASP) & 100.00 & 100.00 & 99.72 \\
& Name-only retrieval & 54.00 & 96.40 & 77.10 \\
& Variable-only retrieval & 61.60 & 80.80 & 81.25 \\
& Structure-only retrieval & 23.00 & 46.00 & 64.25 \\
& Semantic-only retrieval & 0.80 & 2.80 & 50.30 \\
\midrule
Embedding
& w/o fixed transformation rule set & -- & -- & 31.33 \\
\midrule
Extraction
& Direct bit prediction without original code & -- & -- & 52.01 \\
\bottomrule
\end{tabular}}
\vspace{-3ex}
\end{table}

\section{Conclusion}
We present CLASP, an LLM-assisted framework for source code watermarking that reformulates watermark embedding and extraction as semantic reasoning and retrieval-grounded verification. Unlike prior methods that rely on handcrafted rules, parser-/AST-dependent rewriting, or retraining, CLASP enables automatic watermarking without model fine-tuning or language-specific engineering. Across extensive experiments, CLASP demonstrates strong robustness, effectiveness, generalization, training-free deployment, plug-and-play usability, and high capacity, yielding a favorable overall trade-off among these desirable properties. These results show that CLASP provides a practical and broadly applicable solution for source code watermarking.

\bibliographystyle{ACM-Reference-Format}
\bibliography{sample-base}

\end{document}